\documentclass[sigconf]{acmart}
\usepackage{multicol}
\usepackage{multirow}
\usepackage{subfigure}
\usepackage{caption}
\usepackage{listings}
\usepackage{xcolor}
\usepackage[pscoord]{eso-pic}
\newcommand{\placetextbox}[3]{
  \setbox0=\hbox{#3}
  \AddToShipoutPictureFG*{
    \put(\LenToUnit{#1\paperwidth},\LenToUnit{#2\paperheight}){\vtop{{\null}\makebox[0pt][c]{#3}}}%
  }%
}%

\definecolor{codegreen}{rgb}{0,0.6,0}
\definecolor{codegray}{rgb}{0.5,0.5,0.5}
\definecolor{codepurple}{rgb}{0.58,0,0.82}
\definecolor{backcolour}{rgb}{0.95,0.95,0.92}

\lstdefinestyle{mystyle}{
    backgroundcolor=\color{backcolour},   
    commentstyle=\color{codegreen},
    keywordstyle=\color{magenta},
    numberstyle=\tiny\color{codegray},
    stringstyle=\color{codepurple},
    basicstyle=\ttfamily\footnotesize,
    breakatwhitespace=false,         
    breaklines=true,                 
    captionpos=b,                    
    keepspaces=true,                 
    numbers=left,                    
    numbersep=5pt,                  
    showspaces=false,                
    showstringspaces=false,
    showtabs=false,                  
    tabsize=2
}

\AtBeginDocument{%
  \providecommand\BibTeX{{%
    \normalfont B\kern-0.5em{\scshape i\kern-0.25em b}\kern-0.8em\TeX}}}

\setcopyright{none}

\begin{document}

\title{A multithread AES accelerator for Cyber-Physical Systems}

\author{Francesco Ratto}
\email{francesco.ratto@unica.it}
\orcid{0000-0001-5756-5879}
\affiliation{%
  \institution{Università degli Studi di Cagliari}
  \city{Cagliari}
  \country{IT}
}

\author{Luigi Raffo}
\email{luigi.raffo@unica.it}
\affiliation{%
  \institution{Università degli Studi di Cagliari}
  \city{Cagliari}
  \country{IT}
}

\author{Francesca Palumbo}
\email{fpalumbo@uniss.it}
\affiliation{%
  \institution{Università degli Studi di Sassari}
  \city{Sassari}
  \country{IT}
}

\placetextbox{0.5}{1}{\texttt{\small This is the author's version of the work. It is posted here for your personal use. Not for redistribution.}}
\placetextbox{0.5}{0.99}{\texttt{\small The definitive Version of Record was published in }}%
\placetextbox{0.5}{0.98}{\texttt{\textit{\small 20th ACM International Conference on Computing Frontiers (CF ’23), May 9–11, 2023, Bologna, Italy.}}}
\placetextbox{0.5}{0.97}{\texttt{\small http://dx.doi.org/10.1145/3587135.3592819}}%

\renewcommand{\shortauthors}{Ratto F., Raffo L. et Palumbo F.}

\begin{abstract}
Computing elements of CPSs must be flexible to ensure interoperability; and adaptive to cope with the evolving internal and external state, such as battery level and critical tasks. Cryptography is a common task needed in CPSs to guarantee private communication among different devices. In this work, we propose a reconfigurable FPGA accelerator for AES workloads with different key lengths. The accelerator architecture exploits tagged-dataflow models to support the concurrent execution of multiple threads on the same accelerator. This solution demonstrates to be more resource- and energy-efficient than a set of non-reconfigurable accelerators while keeping high performance and flexibility of execution.
\end{abstract}

\keywords{reconfigurable computing, cyber-physical systems, cryptography}

\received{March 2023}

\maketitle


\section{Introduction}
\label{sec::intro}
Cyber-physical systems (CPS) are systems that integrate physical processes with computation and communication. CPS employ various technologies, such as sensors, actuators, and embedded systems, to monitor and control physical processes \cite{Lee08a}. They are meant to act autonomously and react dynamically to different internal and external stimuli, which may affect functional (e.g. executed mission, parameters of the executed algorithms, etc.) and non-functional (e.g. QoS, security degree, energy, etc.) requirements, putting in place different adaptive mechanisms \cite{PalumboFSRMDMPR19}. Security is among the key matters in any modern system including CPS\cite{ChenN22}. Indeed, as CPS become more interconnected and integrated with the internet, they become vulnerable to cyber threats, which can cause serious damage to the physical systems they control, as well as to humans.

Implementing secure CPS comes at a cost. Cryptographic strength, implementation cost, execution speed, and energy consumption can be seen as trade-offs in the implementation of cryptographic systems \cite{4358705}.  The target scenario determines their relative priority. In CPS, where constraints are expected to change, flexibility support is a desideratum also with respect to security aspects. 
As an example, the Advanced Encryption Standard (AES) supports three different key lengths (128, 192, and 256) with a different associated number of encryption steps (10, 12, and 14 respectively). Supporting multiple standards with the same accelerators has several advantages. It can be useful to obtain the desired level of computational cost and security, depending on the type of attack \cite{Biryukov05}, and to provide interoperability among a larger set of devices and protocols. For instance, the IBM Aspera communication protocol uses AES-128 as default \cite{IBM}, while AWS Encryption SDK recommends using the AES with a 256-bit key \cite{AWS}. However, both products allow developers to choose among the three standard key lengths. This is a strong motivation to go for an adaptable solution capable of surfing at runtime among different algorithms, thus allowing concurrent communication with multiple different systems.


Field Programmable Gate Array (FPGA) technologies are gaining momentum in many different domains. They provide not only performance, but also characteristics for coping with rapidly evolving requirements. For example, their intrinsic long-term flexibility is the main reason for automotive embedded-system engineers to adopt FPGAs as a way to keep pace with the evolving requirements in the Autonomous Driving industry \cite{FPGACars}. Similar reasons are making them popular also in the CPS and Internet of Things domains~\cite{magyari2022review}: providing flexibility, at no performance nor energy consumption costs.

In this paper we make use of the tagged-token approach proposed in~\cite{RattoESRP22} for multithread execution and we extend the Multi-Dataflow Composer tool \cite{SauFRRP21} to support it as described in Sect.~\ref{ssec::design-meth}. Then we apply the proposed methodology to the design of a reconfigurable AES accelerator (Sect.~\ref{ssec::aes-acc}). The accelerator is assessed against a set of single-thread designs to show the advantages of hardware multithreading (Sect.~\ref{sec::assessment}). Before drawing the conclusion, a comparison with state-of-the-art solutions is provided and discussed in Sect.~\ref{sec::related}.


\section{Background}  \label{sec::background}
In this section, we first give a brief overview of the AES algorithm (Sect.~\ref{ssec::aes-alg}), which will ease the discussion on its FPGA implementation. Then, in Sect.~\ref{ssec::cgr-mdc}, we introduce the concept of Coarse-Grained reconfigurability and we describe the Multi-Dataflow Composer tool (MDC), which is then used to design the proposed accelerator.

\subsection{AES algorithm}  \label{ssec::aes-alg}
AES \cite{dworkin2001advanced} is a widely used cryptographic algorithm that has been adopted as a standard by the U.S. government and is used to protect sensitive data in various applications, including cloud applications \cite{IBM}, e-commerce \cite{Ecomm}, and disk encryption \cite{Iphone}. The AES algorithm is a symmetric encryption algorithm, which means that the same secret key is used for both encryption and decryption. It operates on blocks of plaintext, typically 128 bits in size, and uses a key of either 128, 192, or 256 bits in length. The algorithm consists of several rounds of substitution and permutation operations, which scramble the plaintext to compute the ciphertext. 

The number of rounds in the AES encryption process depends on the key size: 10 rounds for a 128-bit key, 12 rounds for a 192-bit key, and 14 rounds for a 256-bit key. Each round consists of four main operations: SubBytes, ShiftRows, MixColumns, and AddRoundKey.

\begin{itemize}
    \item SubBytes: In this operation, each byte in the block is replaced with a corresponding byte from a fixed substitution table called the S-box. 
    \item ShiftRows: In this operation, the rows of the block are shifted by a certain number of bytes.
    \item MixColumns: In this operation, each column of the block is multiplied by a fixed matrix.
    \item AddRoundKey: In this operation, the block is XORed with a round key derived from the original secret key. Each round key is unique to a specific round of encryption and is generated from the original secret key using a key expansion algorithm.
\end{itemize}

The AES key expansion algorithm is used to derive a set of round keys from the original secret key. The exact operations depend on the key size and differ for each round. These round keys are used in each round of the AES encryption process to XOR with the state matrix during the AddRoundKey operation. The steps of the algorithm are shown in List.~\ref{list::aes-code}.

\begin{lstlisting}[language=Octave, caption=Pseudo code of the AES algorithm. An iteration of the for loop in called a \emph{round}., label=list::aes-code]
function AES_cipher(plaintext,  exp_key):
  chipertext = plaintext
  AddRoundKey(chipertext, exp_key[0])

  # Each iteration of the loop is a round
  for i_round = 1 step 1 to Nr-1
    SubBytes(chipertext)
    ShiftRows(chipertext)
    MixColumns(chipertext)
    AddRoundKey(chipertext, exp_key[i_round])
  end for
   
  SubBytes(chipertext)
  ShiftRows(chipertext)
  AddRoundKey(chipertext, exp_key[i_round + 1])
   
  return chipertext
\end{lstlisting}

\subsection{Coarse-grained reconfiguration and MDC} \label{ssec::cgr-mdc}
Reconfigurable architectures are meshes of processing elements (PEs) whose functionality and connections can be configured at run time. Depending on the granularity of the PEs, it is possible to have fine-grained (FG) or coarse-grained (CG) reconfigurability \cite{FanniRSSPRT18, tan2021aurora, brandalero2018approximate}. The former, typical of FPGAs, involves bit-level PEs. Resulting in higher flexibility, bit-level programmability is characterized by a configuration time overhead (due to the configuration bitstream size). CG reconfigurability provides word-level PEs, thus achieving less flexibility while guaranteeing faster configuration phases. 
Resource sharing in CG reconfigurable systems leads to area-efficient designs while maintaining the desired set of functionalities available.

\begin{figure*}[t]
    \centering
    \includegraphics[trim=4.2cm 5.2cm 5.5cm 6cm, clip,width=.8\textwidth]{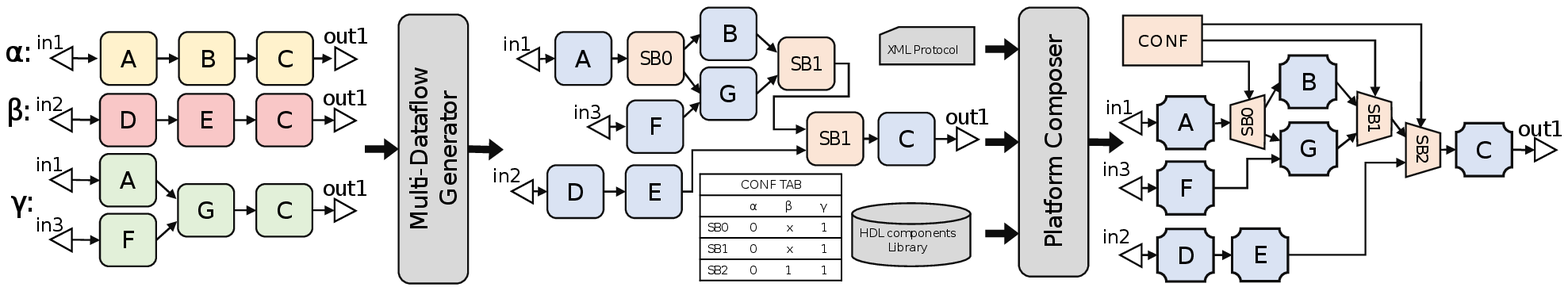}
    \caption{Overview of the MDC functionality, inputs, and outputs. The three input dataflow ($\alpha$, $\beta$, $\gamma$) are merged by the \textit{Multi-Dataflow Generator} in a unique dataflow that shares common actors (\textit{A}, \textit{C}) and addresses tokens depending on the configuration using switching boxes\protect\footnotemark (\textit{SB}).  \textit{Platform-Composer} maps this dataflow in a hardware accelerator using the provided actor library (\textit{HDL components library}) and communication protocol (\textit{XML protocol}). }
    \label{fig::mdc}
\end{figure*}

MDC is an open-source tool for designing and deploying CG reconfigurable accelerators\footnote{https://github.com/mdc-suite/mdc}. It takes as input applications specified as dataflows: they can be different applications with common processing steps (common actors), to obtain functional reconfiguration or different versions of the same application, to obtain non-functional reconfiguration. The dataflows corresponding to such applications are combined together, and the resulting HDL top module is automatically generated (see Fig.~\ref{fig::mdc}).

\section{Proposed multithread architecture}  \label{sec::prop-arch}
In this section, we describe the general design methodology for concurrent multithreading (Sect.~\ref{ssec::design-meth}) and we apply it to the AES application (Sect.~\ref{ssec::aes-acc}).

\subsection{Design methodology}   \label{ssec::design-meth}
In ~\cite{RattoESRP22} the authors proposed a model-based design approach that takes as initial model the single-thread dataflow specification of an application. Without an explicit need for data synchronization, it allows designing a corresponding multithread hardware accelerator. 
Hardware accelerators can be derived from dataflows mapping each actor directly into a module, while FIFOs' and tokens' flow ensures the execution correctness without the need for centralized control.

To support hardware multithreading a tag is added to each token. The tag indicates which thread the token belongs to, and so it allows the differentiation of tokens from different threads. Once tokens are tagged, FIFOs and actors must meet a set of requirements to ensure the correct execution of each thread:
\begin{enumerate}
    \item \label{req:output_tag} A firing actor has to tag the output tokens with the same tag of the input ones.
    \item \label{req:matching_tag} The firing rules have to be adjusted so that only tokens belonging to the same thread are able to fire the execution.
    \item \label{req:out-of-order_read} FIFOs must provide a semi-out-of-order read, letting the reading actors choose among the first token of each flow of execution. This is necessary to prevent deadlocks.
\end{enumerate}

An augmented FIFO interface is adopted to allow the reading actor to choose from which thread it wants to read from and also to know the status of the FIFO with respect to that particular thread. For the FIFOs, it is adopted an architecture with two blocks of memory, one for storing tokens and one for their order. Actors process tokens of different threads in the same computation logic.

One of the contributions of the proposed work is the MDC-tool extension to support multiple concurrent threads. The set of possible applications is defined at design time: a dataflow description corresponding to each application must be provided to the tool. To support multithreading, the tool has been extended with a new feature. Depending on the number of concurrent threads of execution that the accelerator must support, the interface is set to have an adequate number of bits for the tag and for the control signals. Moreover, the sboxes, which are in charge of addressing tokens, have been designed to match the token tag with the selected configuration for that thread.

\begin{figure}[tb]
    \centering
    \includegraphics[trim=8cm 6.5cm 9cm 6cm, clip,width=\linewidth]{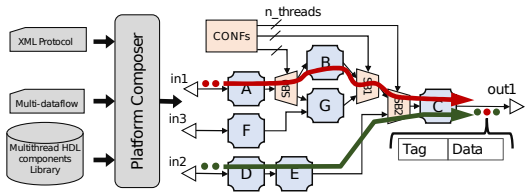}
    \caption{The output of the extended \emph{Platform Composer} is depicted. Two threads, depicted in red and green, are processed in two different configurations. The \emph{XML protocol} file is used to specify the number of concurrent threads and describe the new interface. The \emph{sboxes} address tokens considering both the configuration and their tag.}
    \label{fig::mdc-mt}
\end{figure}

In the example shown in Fig.~\ref{fig::mdc-mt} two concurrent threads are running, one executing the $\alpha$ dataflow and one the $\beta$ dataflow. The HDL actors provided to the Platform Composer are able to process tokens including their tags. The maximum number of concurrent threads and the corresponding tag width is specified through the XML protocol file. The tokens of different threads, depicted with different colors, are addressed along the two highlighted paths.

\footnotetext{The SBs used by MDC to address tokens shall not be confused with the S-box used during AES computation. As their name was already set in previous works, we preferred not to change it.}

\subsection{AES multithread accelerator} \label{ssec::aes-acc}
\begin{figure}[tb!]
    \centering
    \subfigure[AES-128 dataflow]
    {
        \includegraphics[trim=1.1cm 4cm 15.7cm 2cm, clip,width=0.7\linewidth]{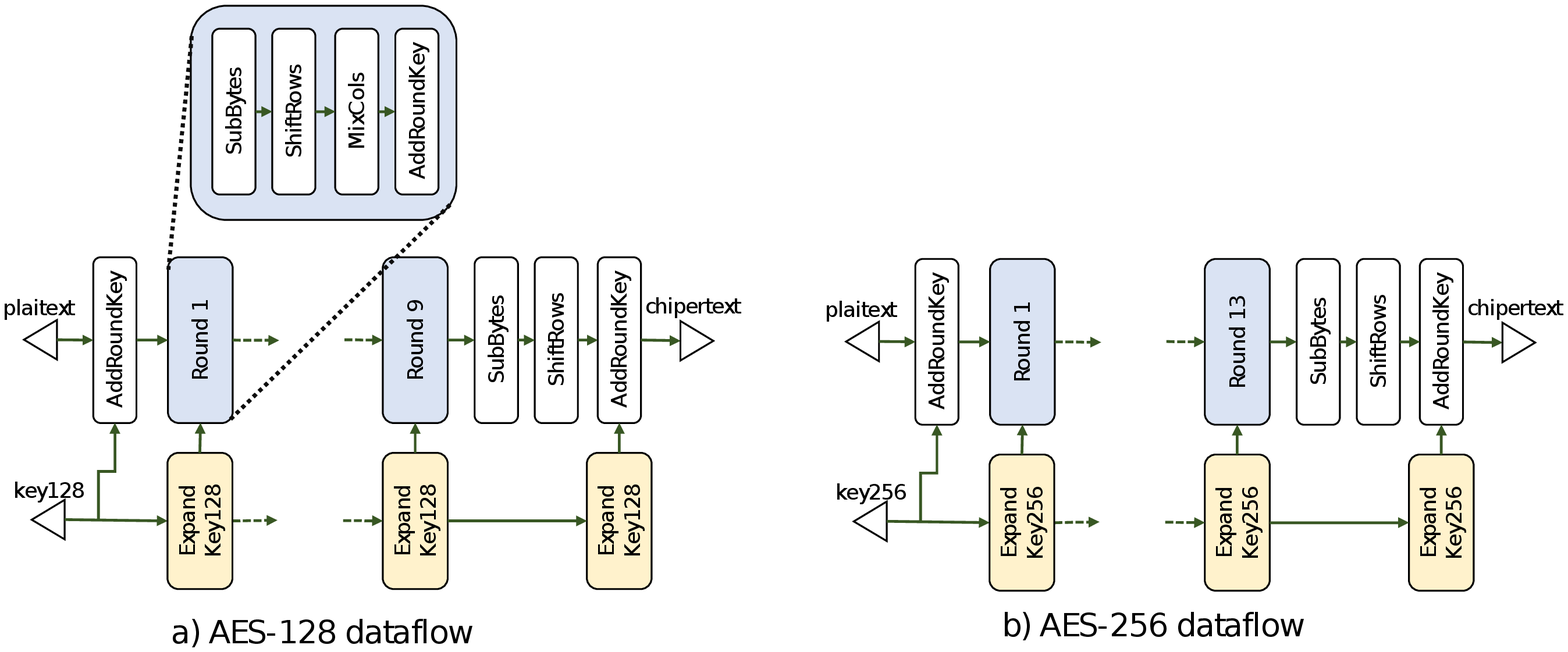}
        \label{fig::aes-128-df}
    }
    \subfigure[AES-256 dataflow]
    {
        \includegraphics[trim=15.4cm 4cm 1.4cm 6.5cm, clip,width=0.7\linewidth]{figs/aes-df.eps}
        \label{fig::aes-256-df}
    }
    \caption{Dataflow descriptions of AES-128 and AES-256 used to derive the accelerators. In Fig.~\ref{fig::aes-128-df} the inner pipeline of a round is depicted.}
    \label{fig::aes-df}
\end{figure}

In this work, we intend to demonstrate that the multithreading approach can be effectively adopted to support flexible security profiles by selecting the AES-128 and AES-256 as the two applications to be accelerated.
The block encrypted with AES is made of 128 bits in both cases. Typically the amount of data that needs to be encrypted is larger, so the accelerator must provide high throughput to encrypt the input data effectively. FPGAs are an ideal technology for achieving that goal, as they allow unrolling the rounds in a deep pipeline. The same holds for the key expansion steps which are associated with each round. The resulting dataflow describing the unrolled rounds is shown in Fig.~\ref{fig::aes-df}. Coherently with the AES standard, the actor implementing the rounds is the same in AES-128 and AES-256 dataflows, while the key expansion stages are different for the two algorithms. In addition, the architecture of a round actor has been designed with an inner pipeline, where each stage corresponds to a step of a round. This is done, again, to achieve high throughput and avoid long critical paths. In the proposed architecture we support key expansion to be executed by the hardware accelerator. This choice is particularly effective with respect to a software expansion when the key is changed frequently.

We used MDC to automatically merge these two dataflows and the newly developed feature of the Platform Composer to deploy it as a multithread accelerator. The accelerator is capable of computing two threads concurrently, each one being either AES-128 or AES-256. The configuration associated with each thread can be changed at runtime just by writing the configuration ID in the associated register. It is worth noticing that the same methodology could have been used to merge any other set of AES-like applications that use a different key length and/or a different number of rounds.

\section{Assessment} \label{sec::assessment}
The proposed accelerator has been tested with the Vivado Xilinx tool suite targeting the ZCU102 evaluation board. The experimental setup and the outcomes are described in Sect.~\ref{ssec::setup}~and~\ref{ssec::results} respectively. Finally, a detailed exploration on the energy consumption of the design is presented in Sect.~\ref{ssec::energy}.

\subsection{Setup} \label{ssec::setup}
To test the advantage of using CG reconfigurability and multithreading, we considered the following designs:
\begin{itemize}
\item \emph{AES-128-ST}: an AES-128 single thread accelerator obtained from the dataflow shown in Fig.~\ref{fig::aes-128-df};
\item \emph{AES-256-ST}: an AES-256 single thread accelerator obtained from the dataflow shown in Fig.~\ref{fig::aes-256-df};
\item \emph{AES-PARALL}: a design without resource sharing. The two previous accelerators are instantiated in parallel.
\item \emph{AES-RECONF}: a multithread CG reconfigurable AES accelerator as described in Sect.~\ref{ssec::aes-acc}.
\end{itemize}

To compare these designs we considered the following metrics:
\begin{itemize}
\item \emph{CLBs}: (Configurable Resource Blocks) it is the number of slices needed to implement a design. It is obtained from the logic synthesis carried out with the Vivado Xilinx tool;
\item \emph{Latency}: it is the time needed by an accelerator to process an AES block of 128 bits; 
\item \emph{Throughput}: it is the rate at which an accelerator can process AES blocks of 128 bits;
\item \emph{Power}: it is the average power dissipated during the processing of a given workload. It is estimated by the Vivado Power Analyzer considering the switching activity of a post-synthesis simulation; 
\end{itemize}

\subsection{Results} \label{ssec::results}
The results obtained for the 4 designs are reported in Tab.~\ref{tab::results}.
The column CLBs shows that resource sharing among configurations can lead to a significant saving of 22,6\% while guaranteeing the same functionalities provided by two distinct accelerators.
The latency of encrypting a block in the multithread AES-RECONF accelerator remains the same as in the single-thread accelerators, as the block has to go through the same stages of the pipeline.
The same holds for the throughput which is set by the rounds pipeline. The only difference here is that two accelerators in parallel can possibly double the throughput but they do not improve the block latency.

\begin{table}[h]
\renewcommand\arraystretch{1.1} 
\centering
\caption{Results in terms of resources and performance of the proposed designs. The \% line reports the percentage variation of the AES-RECONF design with respect to the AES-PARALL one.}
\begin{tabular}{l c c c c c c } 
\toprule
\multirow{2}{1.5cm}{Design}     &CLBs &Latency &Throuput \\
                                &[\#] &[ns] &[Gbps] \\ \midrule
AES-128-ST	    &23556	&250	      &25\\
AES-256-ST	    &35582	&350	       &25\\
AES-PARALL  &59138	&250/350	    &50\\
AES-RECONF	&45792	&250/350	   &25\\                                                                        
\%	    &-22,6\%	 &0\%	&-50\%\\ \bottomrule               

\end{tabular}
\label{tab::results}
\end{table}

Power consumption is strictly correlated to the adopted testing scenario. For this reason, we evaluated the power consumption of each design with five test cases: \begin{enumerate}
    \item \emph{idle} -  the accelerator is not processing any data, but still, it dissipates some power due to the clock activity;
    \item \emph{low} - the accelerator processes only one block, so the pipeline is not fully active;
    \item \emph{high 128} - the accelerator encrypts 100 blocks with AES-128 algorithm, which are enough to completely fill the pipeline;
    \item \emph{high 256} - the accelerator encrypts 100 blocks with AES-256;
    \item \emph{high both} - the accelerator encrypts concurrently 100 blocks with AES-128 algorithm and 100 blocks with AES-256 algorithm. 
\end{enumerate}
The results of this exploration are shown in Tab.~\ref{tab::power}. The power values for the AES-PARALL design are obtained as the sum of the two single designs. When only one of the two algorithms is executed, the other non-active design  consumes as in \emph{idle}.
The AES-RECONF shows a reduced power consumption in all scenarios excluding the \emph{high 256}, where a small increase (1\%) is there, compared with the AES-PARALLEL. In the first 4 scenarios, the power reduction is obtained at no cost in terms of performance, as the AES-RECONF achieves the same throughput and latency as the AES-PARALL. In the \emph{high both} tests a significant power saving (36\%) is obtained at the cost of a reduced throughput, as shown in Tab.~\ref{tab::results}. 

\begin{table}[h]
\renewcommand\arraystretch{1.1}
\centering
\caption{Power dissipation in the proposed test cases. The \% line reports the percentage variation of the AES-RECONF design with respect to the AES-PARALL one.}
\begin{tabular}{l c c c c c}   
\toprule
\multirow{2}{*}{Design}		&\multicolumn{5}{c}{Power consumption [mW]}\\
 &\emph{idle}	&\emph{low}	&\emph{high 128}	&\emph{high 256}	&\emph{high both} \\ \midrule
AES-128-ST		&40	&57	&241	&-	&- \\
AES-256-ST		&71	&93	&-	&269	&- \\
AES-PARALL		&111	&150	&312	&309	&510 \\
AES-RECONF		&102	&135	&293	&313	&376 \\
\%		&-9\%	&-11\%	&-6\%	&1\%	&-36\% \\ \bottomrule
\end{tabular}
\label{tab::power}
\end{table}

\subsection{Energy profiles} \label{ssec::energy}
In the context of CPS energy consumption is more of a concern than power itself. Considering the power consumption data presented in Tab.~\ref{tab::power}, it is possible to evaluate the impact on energy consumption in a set of scenarios. For this analysis, we consider the AES-PARALL and AES-RECONF designs, powered by a battery of 5000mAh. The testing scenarios are reported in Tab.~\ref{tab::scenarios}, where each power profile is associated with a corresponding rate of usage. To restrict the set of possibilities, in all of the scenarios rates of \emph{high 128},	\emph{high 256}, and \emph{high both} are equal. In the first, the rate of \emph{low} is fixed at 0.25, and the rate of \emph{idle} decreases at steps of 0.25. In the second group, the rate of \emph{idle} is fixed at 0, and the rate of \emph{low} decreases at steps of 0.25.

\begin{table}[h]
\renewcommand\arraystretch{1}
\centering
\caption{Rate of usage of each condition in 8 different scenarios.}
\begin{tabular}{l c c c c c}   
\toprule
ID   &\emph{idle}	&\emph{low}	&\emph{high 128}	&\emph{high 256}	&\emph{high both} \\ \midrule
1    &0.75	&0.25	&0	&0	&0\\
2    &0.5	&0.25	&0.08	&0.08	&0.08\\
3    &0.25	&0.25	&0.17	&0.17	&0.17\\
4    &0	&0.25	&0.25	&0.25	&0.25\\ \midrule
5    &0	&0.75	&0.08	&0.08	&0.08\\
6    &0	&0.5	&0.17	&0.17	&0.17\\
7    &0	&0.25	&0.25	&0.25	&0.25\\
8   &0	&0	&0.33	&0.33	&0.33\\
\bottomrule
\end{tabular}
\label{tab::scenarios}
\end{table}

\begin{figure}[tb]
    \centering
    \includegraphics[trim=0cm 0cm 0cm 0cm, clip,width=\linewidth]{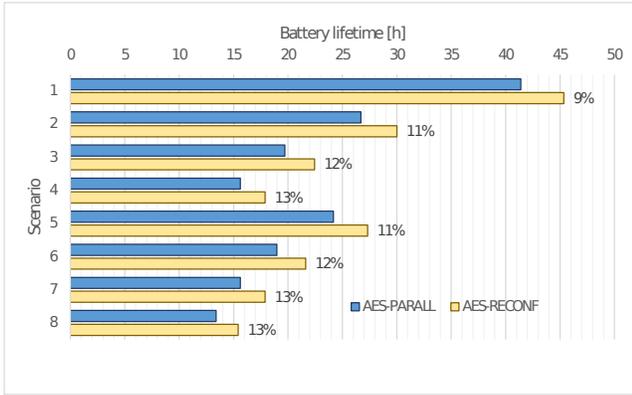}
    \caption{Duration of a 5000mAh battery considering the scenarios described in Tab.~\ref{tab::scenarios}. The number above the AES-RECONF columns reports the average variation with respect to the corresponding AES-PARALL column. }
    \label{fig::battery-lifetime}
\end{figure}

The power consumption data from Tab.~\ref{tab::power} are used within the scenarios described in Tab.~\ref{tab::scenarios} to derive the consequent energy consumption. This is then exploited in Fig.~\ref{fig::battery-lifetime} where the predicted battery lifetime is depicted. We can see that in all cases the AES-RECONF design leads to a longer battery duration.

\section{Related works}
\label{sec::related}
The flexibility of the FPGAs substrate offers the possibility to apply a large set of techniques to improve the design of an accelerator for a given application. 

In \cite{shahbazi2020high} the authors proposed a new affine transformation to reduce the area of Sub-Bytes, which is the combination of inverse isomorphic and affine transformations. The AES rounds are unrolled to achieve high throughput. 

El Maraghy et al. \cite{el2013real} adopted an iterative looping method multistage sub-pipelining architecture to achieve area efficiency, while the software application is linked to the hardware design through the Xilinx MicroBlaze soft processor core.

In \cite{chen2019high} the authors  use the deep pipeline and full expansion technology to implement the AES encryption algorithm on FPGA. Their goal is to improve the encryption speed of big-data workloads using High-level Synthesis.

Visconti et al. \cite{visconti2021high} present a fast and lightweight AES-128 cipher based on the Xilinx ZCU102 FPGA board, suitable for 5G communications. Both encryption and decryption algorithms have been developed using a pipelined approach.

Tab.~\ref{tab::related} shows a comparison among the aforementioned literature works and the one presented in this paper (last two rows). To have a more fair comparison, we included in Tab.~\ref{tab::related} also the AES-128-ST accelerator described in Sect.~\ref{ssec::setup}. This latter obtains state-of-the-art performance in terms of latency and throughput. Only the design proposed in \cite{shahbazi2020high} achieves a better throughput thanks to the SubByte optimization. The design presented in \cite{el2013real} achieves lower latency at the cost of considerably lower throughput. 

Considering this paper's contribution, the AES-RECONF is capable of offering the same level of performance as the AES-128-ST. The latency will be larger when implementing the AES-256 standard due to the larger number of rounds. These results and considerations prove the effectiveness of the general methodology presented in Sect.~\ref{ssec::design-meth} when implementing an adaptable AES. Indeed, the proposed flexible solution will come at no relevant implementation costs, making this approach particularly suitable for adoption in CPS environments.

\begin{table*}[h]
\renewcommand\arraystretch{1.2} 
\centering
\caption{Related work table. For each accelerator, it is reported the supported standard, whether the key expansion can be performed in hardware (Key exp. column), and its performance in terms of area (CLBs column), latency, and throughput.}
\begin{tabular}{l c c c c c c c }      
\toprule
						
		&\multirow{2}{*}{Year} &\multirow{2}{*}{Target} &\multirow{2}{*}{Standard} &Key 	&CLBs	&Latency	&Throughput\\
 	&  & & &exp.	&[\#]	&[ns]	&[Gbps] \\
\midrule
ElMaraghy \cite{el2013real}  &2013  &Virtex-5   &AES-128        & YES	& 303	& 96	& 1,33\\
Chen \cite{chen2019high}    &2019 &Kintex-7  &AES-128		        & NO	& 12640	& 250	& 31,29\\
Shahbazi \cite{shahbazi2020high}  &2020 &Virtex-5   &AES-128		           & NO	& 5974	& -	& 79,51\\
Visconti \cite{visconti2021high}    &2021  &ZCU102 &AES-128		       & YES	& 1631	& -	& 28\\
AES-128-ST   &2023	&ZCU102     &AES-128	     & YES	& 18584	& 166	& 32,25\\ \midrule
\multirow{2}{*}{AES-RECONF}   &\multirow{2}{*}{2023}  &\multirow{2}{*}{ZCU102}   &AES-128		&\multirow{2}{*}{YES}	&\multirow{2}{*}{43401}	& 166	&\multirow{2}{*}{32,25}\\
                            & & &AES-256       &                        &                       &233   & \\
\bottomrule               

\end{tabular}
\label{tab::related}
\end{table*}
\section{Conclusions}
CPS are vulnerable systems, which require integrating cryptography support to mitigate cyber threats. The flexibility of execution is key for providing interoperability and exploiting trade-offs. In such a context the AES chiper is an example of a widely adopted standard that supports different configurations through different key lengths. The more the supported configurations are, the higher the impact in terms of resource utilization. Hardware multithreading matched with resource sharing helps in realizing a compromise between flexibility and performance.

The more supported configurations, the higher the impact in terms of resource utilization. The support of hardware multithreading helps in realizing a compromise between flexibility and performance.

In this work, we extended the MDC tool with a novel back-end feature to support hardware multithreading through token tagging and we demonstrated that it is possible to leverage it to obtain the desired flexibility and reduced energy consumption.  Our adaptable AES accelerator implementation demonstrated to be able to guarantee the same performance of a non-reconfigurable state-of-the-art approach while featuring different AES algorithms over the same FPGA platform.

\begin{acks}
Prof. Francesca Palumbo would like to thank EU Commission for supporting her work through the SECURED project, funded within the EU Horizon Europe initiative under grant number 10109571.
\end{acks}

\bibliographystyle{ACM-Reference-Format}
\bibliography{sample-base}

\end{document}